\documentclass[preprint,12pt]{elsarticle}

\usepackage{amssymb}
\usepackage{amsmath}

\journal{Journal of Electrostatics}

\usepackage{times}
\usepackage{graphicx}
\usepackage{amsmath} 
\usepackage{amssymb}
\usepackage[runin]{abstract}
\usepackage{sectsty}
\usepackage{anyfontsize}
\usepackage{fancyhdr}
\usepackage{indentfirst} 
\usepackage{booktabs} 
\usepackage{comment}
\usepackage{overpic}

\begin{document}

\begin{frontmatter}
\title{Energy analysis of 2D electro-thermo-hydrodynamic turbulent convection}

\author[label1]{Owen Hutchinson} 
\author[label1]{Katerina Kostova} 
\author[label2]{Jian Wu} 
\author[label1]{Yifei Guan\fnref{label3}} 
\fntext[label3]{Corresponding author: guany@union.edu}

\affiliation[label1]{organization={Deparment of Mechanical Engineering},
            addressline={Union College}, 
            city={Schenectady},
            postcode={12308}, 
            state={NY},
            country={USA}}
\affiliation[label2]{organization={School of Energy Science and Engineering},
            addressline={Harbin Institute of Technology}, 
            city={Harbin},
            postcode={150001}, 
            country={China}}

\end{frontmatter}
\begin{abstract}
    Turbulent convection is ubiquitous in fluid systems. In particular, multi-physical convection problems involve mass, heat, and particle transfer. When the particles are charged and driven by a high-voltage electric field, both buoyancy and electric forces contribute to driving and maintaining the convection. In this work, we perform numerical analysis using a high-fidelity Fourier-Chebyshev spectral solver. We further derive the dynamical systems governing the kinetic energy, the enstrophy, the potential energy, and the electric energy analytically. Using the simulated data, we apply a long short-term memory recurrent neural network to predict the chaotic time series of domain-average energy terms. Finally, we perform a data-driven modal decomposition to show the coherent structures that contain energy and enstrophy in 2D turbulent convection. 
\end{abstract}

\section{Introduction}

Electro-hydrodynamics (EHD) is a multi-physical discipline that describes the three-way interactions between the charged particles, neutral molecules, and the electric field on a macroscopic scale, which involves fluid mechanics and electrodynamics~\cite{felici1978electroconvection,castellanos2014electrohydrodynamics, wang2021lattice,wu2015two,zhang2015characteristics,zhang2015modal,he2022moffatt, liu2020scaling,liu2022critical}. Additionally, electro-thermo-hydrodynamics (ETHD) involves one more complexity, the heat transfer or temperature field~\cite{he2021flow,ma2021electro,liu2025review}. ETHD convection has been a research focus since its first introduction to model the geo-convection of the mantle~\cite{gross1966electrically,pontiga1992onset}. ETHD and EHD have since been analyzed in multiple backgrounds, e.g., solid–liquid interface~\cite{he2021numerical}, Taylor cone~\cite{wang2021massively}, and enhanced heat transfer~\cite{ma2022experimental}. In recent years, many studies have focused on the linear stability analysis of the ETHD systems~\cite{guan2021monotonic,he2021flow} and transition routes to chaos~\cite{li2020transition,wang2022transition}. Despite the broad interest in the stability analysis of laminar or mildly chaotic ETHD convection, only a few studies have attempted to analyze the turbulent and highly chaotic (high Reyleigh number, $Ra$, and high electric Rayleigh number, $T$) regimes~\cite{kourmatzis2012turbulent,guan2025numerical}. In particular, the recent paper of our group presents the relationship between the potential energy and electric energy budgets and the kinetic energy dissipation~\cite{guan2025numerical}.

The numerical study of ETHD convection flows is routinely performed with high-fidelity numerical simulations~\cite{castellanos1987numerical,chicon1997numerical,traore2012two,wu2013onset,luo2016lattice,luo2018hexagonal,guan2019two,guan2019numerical,wang2021lattice,guan2025numerical}, resulting in a large amount of high-resolution spatio-temporal data.  
In spite of the high-dimensional nature of turbulent (highly chaotic) ETHD convection, many phenomena in electro-thermo-convection and other similar systems can be characterized by a few dominant coherent structures~\cite{holmes2012turbulence,Taira2017aiaa,taira2020aiaa,brunton2019data}, or \emph{modes} as in modal analysis.  
High-fidelity simulations (e.g., direct numerical simulations or DNS) often obscure this simplicity, making optimization and control tasks prohibitively expensive~\cite{Brunton2020arfm}. 
Therefore, there is a need for accurate and efficient reduced-order models that describe ETHD turbulence, similar to the Lorenz model for Rayleigh-B\'enard convection (RBC)~\cite{lorenz1963deterministic}. 
The Lorenz model was obtained by Galerkin projection of the governing equations onto a reduced set of three Fourier modes that was earlier proposed by Saltzman~\cite{saltzman1962finite}. 
Instead of extracting these modes manually (analytically), modes can also be extracted automatically (from data), for example via proper orthogonal decomposition (POD)~\cite{lumley1967structure,berkooz1993proper,lumley2007stochastic,holmes2012turbulence,brunton2019data} or dynamic mode decompostion (DMD)~\cite{brunton2019data,kutz2016dynamic}. Modal analysis also facilitates other practice such as flow control~\cite{ito1998reduced,noack2011reduced,mohan2018deep,chao2023hybrid,xi2023improved,alawi2024artificial,cunegatto2024multi} and fluid-structure interaction analysis~\cite{taira2017modal,taira2020aiaa}.

In order to predict the temporal evolution of chaotic dynamical systems, recent advances focus on data-driven prediction: e.g., artificial neural network~\cite{karunasinghe2006chaotic}, reservoir computing~\cite{chattopadhyay2020data, lu2017reservoir, carroll2018using,nathe2023reservoir}, and recurrent neural network (RNN)~\cite{zhang2000predicting,li2016new}. In particular, the Long Short-Term Memory (LSTM) network is an RNN architecture designed to model temporal dependencies in sequential data. Different from conventional RNNs, which often suffer from vanishing or exploding gradients when learning long-range correlations, LSTMs introduce a gated structure, comprising input, output, and forget gates, that enables the selective retention and updating of information over extended time horizons. This capability makes LSTMs particularly effective for capturing nonlinear and dynamic patterns in time series, making them widely adopted in domains such as speech recognition, natural language processing, and scientific modeling of complex dynamical systems~\cite{graves2012long,vlachas2018data,langeroudi2022fd}.

In the current study, we follow up upon our recent work~\cite{guan2025numerical} to (1) derive the governing equations for kinetic energy, enstrophy, potential energy, and electric energy; (2) train LSTMs to predict the time series of domain-average kinetic energy, potential energy, and electric energy; (3) perform data-driven modal decomposition using the proper orthogonal decomposition (POD); and (4) compare the modal analysis and identify the coherent structures related to energy concentration. 


The remainder of the paper is as follows. Section~\ref{sec:method} presents the dimensionless governing equations of the original 2D ETHD system, and the derivation of the dynamical systems of the energy terms. It further describes the LSTM, 2D wavelet analysis, and POD analysis used in this work. Section~\ref{sec:Results} presents the numerical results. Conclusion and future work discussion are presented in Section~\ref{sec:conclusion}.

\section{Methodology}~\label{sec:method}
\subsection{Governing equations of the ETHD system}
The governing equations for ETHD convection system include the conservation of mass and momentum (Navier-Stokes equations with the electric and buoyancy forcing terms), the conservation of electrical current equation (transport of charge density), the Poisson equation for electric potential with the space-charge effect, and the conservation of potential energy equation (heat transport). The effects of viscous dissipation and the Joule heating on the thermal energy are neglected for simplicity. Following the Boussinesq approximation, the dimensionless governing equations can be written as:~\cite[e.g.,][]{kourmatzis2012turbulent,li2020transition,guan2021stable,guan2025numerical}
\begin{equation}
    \nabla\cdot\mathbf{u}= 0, \label{eq:conMass}
\end{equation}
\begin{equation}
    \frac{\partial{\mathbf{u}}}{\partial t}+(\mathbf{u}\cdot\nabla)\mathbf{u} = -\nabla P+\nu\nabla^2\mathbf{u} + F_Eq\mathbf{E}+F_\theta\theta \mathbf{e}_y,
\end{equation}
\begin{equation}
    \frac{\partial q}{\partial t}+\nabla\cdot[q(\mathbf{u}+\mathbf{E})]=\nu_E\nabla^2q,
\end{equation}
\begin{equation}
    \nabla^2\phi = -Cq/4,
\end{equation}
\begin{equation}
    \mathbf{E}=-\nabla\phi,
\end{equation}
\begin{equation}
    \frac{\partial\theta}{\partial t}+\mathbf{u}\cdot\nabla\theta-u_y=\nu_\theta\nabla^2\theta. \label{eq:Temperature}
\end{equation}

Here, $\mathbf{e}_y=(0,1)$ is the unit vector pointing towards the $y$ direction (opposite of gravity); $\mathbf{u}=(u_x,u_y)$ is the 2D velocity vector; $p$ is the pressure scalar; $\mathbf{E}=(E_x,E_y)$ is the 2D electric field vector; $\theta$ is the temperature anomaly defined as the difference between the temperature and conductive temperature profile ($T_{\text{conduction}}=-y$); $q$ is the net charge density; and $\phi$ is the electric potential. Note that all parameters are dimensionless~\cite{guan2025numerical}.

The system is controlled by the initial and boundary conditions, together with the six dimensionless parameters: $\nu$ is the molecular viscosity; $F_E$ is the electric forcing magnitude; $F_\theta$ is the buoyancy forcing magnitude; $\nu_E$ is the charged particle viscosity; $C$ is the charge injection strength; and $\nu_\theta$ is the thermal diffusivity.

\subsection{Boundary conditions}
The non-slip boundary condition ($\mathbf{u}_{y=\pm h}=0$) are applied to the fluid flow at the upper ($y=h$) and lower ($y=-h$) planes. And the periodic boundary condition is applied to the lateral ($x$) direction. The boundary conditions of the other variables are listed as follows.
\begin{equation}
    \phi_{y=-h} = 1, \: \phi_{y=h} = 0,
\end{equation}
\begin{equation}
    q_{y=-h} = 1,\: (\partial q/\partial y)_{y=h} = 0,
\end{equation}
\begin{equation}
    \theta_{y=\pm h} = 0.
\end{equation}

In this work, we investigate 8 cases with the dimensionless parameters shown in Table~\ref{tab:1}.
\begin{table}[t]
	\caption{\small Dimensionless parameters governing the dynamics of the turbulent ETHD convection}
	\begin{center}
		\begin{tabular}{@{}rrrrc|cc@{}}\toprule\label{tab:1}
            
			Case & $\nu$ & $F_\theta$ & $\nu_\theta$ & $C$ & $\langle E_u\rangle_{V,t}$ & $\langle \Omega\rangle_{V,t}$ \\
			\midrule
            $1$ & $0.050$ & $1.12\times 10^3$ & $7.14\times10^{-3}$ & $0$  & $50.68$ & $1792.90$ \\
            $2$ & $0.050$ & $1.12\times 10^3$ & $7.14\times10^{-3}$ & $10$ & $52.78$ & $1891.20$ \\
            $3$ & $0.033$ & $4.96\times 10^2$ & $4.76\times10^{-3}$ & $10$ & $27.88$ & $978.97$  \\
            $4$ & $0.025$ & $2.79\times 10^2$ & $3.57\times10^{-3}$ & $10$ & $21.85$ & $722.85$  \\
            $5$ & $0.10$  & $6.25\times 10^3$ & $1.00\times10^{-3}$ & $0$  & $49.98$ & $1558.30$ \\
            $6$ & $0.10$  & $6.25\times 10^3$ & $1.00\times10^{-3}$ & $10$ & $51.74$ & $1598.10$ \\
            $7$ & $0.050$ & $1.56\times 10^3$ & $5.00\times10^{-4}$ & $10$ & $16.90$ & $489.90$\\
            $8$ & $0.033$ & $6.94\times 10^2$ & $3.33\times10^{-4}$ & $10$ & $12.34$ & $340.15$ \\	
			\bottomrule
		\end{tabular}
	\end{center}
\end{table}

In addition to these four parameters, the other two dimensionless governing parameters remain constants: $F_E=250$ and $\nu_E=2.00\times10^{-4}$. For Cases 1 and 5 where $C=0$, the ETHD system reduces to the classical Rayleigh-B\'{e}nard Convection (RBC). The dimensionless parameters chosen here cover a broad range: $Pr\in[7,100]$, $Ra\in[5\times10^7,10^9]$, and $T\in[10^3,4\times10^3]$, where $Pr$ is the Prandtl number, $Ra$ is the Rayleigh number, and $T$ is the Taylor number or electric Rayleigh number~\cite{guan2025numerical}.

\subsection{Kinetic, potential, and electric energy and enstrophy equations}
The kinetic energy, $E_u=\frac{1}{2}\mathbf{u}^2=\frac{1}{2}(u_x^2+u_y^2)$, can be obtained by multiplying the momentum equation by the velocity vector $\mathbf{u}$ and domain averaging $\langle \cdot \rangle_V$:
\begin{equation}
    \frac{\partial \langle E_u\rangle_V}{\partial t}= -\langle\epsilon_K\rangle_V+ F_E\langle q\mathbf{u}\cdot\mathbf{E}\rangle_V + F_\theta\langle\theta u_y\rangle_V.
\end{equation}

The enstrophy, $\Omega=\frac{1}{2}\omega^2$, where $\omega = \nabla\times\mathbf{u}$, can be obtained in a similar way:
\begin{equation}
    \frac{\partial \langle \Omega\rangle_V}{\partial t}= -\langle\epsilon_\omega\rangle_V+ F_E\langle \omega \nabla\times(q\mathbf{E})\rangle_V + F_\theta\langle\omega\frac{\partial \theta}{\partial x}\rangle_V.
\end{equation}

Here, $\langle\epsilon_K\rangle_V=\nu\langle\omega^2\rangle_V$, and $\langle\epsilon_\omega\rangle_V=\nu\langle|\nabla\omega|^2\rangle_V$ are energy and enstrophy dissipation terms. 

The potential energy, $E_p=-F_\theta y(\theta-y)$, can be obtained by multiplying the equation for $\theta$ by $y$, followed by a domain averaging:
\begin{equation}
    \frac{\partial \langle E_p\rangle_V}{\partial t}=F_\theta\nu_\theta(N_u-1)-F_\theta \langle \theta u_y\rangle_{V}
\end{equation}

Here, $N_u$ is the Nusselt number that varies in time. The potential energy equation is consistent with the previous studies on Rayleigh-B\'{e}nard Convection~\cite{winters1995available, hughes2013available}.

Similarly, the electric energy, $E_e=F_Eq\phi$, can be obtained by multiplying the equation for $q$ by $\phi$, followed by a domain averaging:
\begin{eqnarray}
    \frac{\partial \langle E_e\rangle_V}{\partial t} &=& F_E\langle p_E\rangle_V -F_E\langle q\mathbf{u}\cdot\mathbf{E}\rangle_V \\
    &-&F_Eq\langle\nabla\phi\cdot\nabla\phi-\frac{\partial \phi}{\partial t}\rangle_V\\
    &-&F_E\nu_E\langle\nabla\phi\cdot\nabla q\rangle_V.
\end{eqnarray}
Here, $\langle p_e\rangle_{V}=\Delta\phi I_0N_e/H$ is the total electric energy injection into the ETHD system across the boundary, where $\Delta\phi$ is the electric potential difference between the lower and upper planes, and $H$ is the height. $I_e = \langle\mathbf{i}\rangle_{A,\:y=-1}$ and $N_e = I_e/I_0$ is the electric Nusselt number. $I_0$ equals $I_e$ at the hydrostatic state~\cite{zhang2024coulomb,guan2025numerical}.

The relation among the kinetic, potential, and electric energy can be summarized as follows.
\begin{eqnarray}
    \frac{\partial\langle E_u\rangle_V}{\partial t} &=& -D_u + \Phi_E + \Phi_\theta\label{eq:eu_vs_t}\\
    \frac{\partial\langle E_p\rangle_V}{\partial t} &=& N_\theta - \Phi_\theta\label{eq:ep_vs_t}\\
    \frac{\partial\langle E_e\rangle_V}{\partial t} &=& N_E - \Phi_E - D_{elec}\label{eq:ee_vs_t}
\end{eqnarray}

Here, $D_u=\langle\epsilon_K\rangle_V$ is the viscous dissipation; $\Phi_E=F_E\langle q\mathbf{u}\cdot\mathbf{E}\rangle_V$ is the electric energy transferred into kinetic energy; $\Phi_\theta=F_\theta\langle \theta u_y\rangle_V$ is the potential energy transferred into kinetic energy; $N_\theta=F_\theta\nu_\theta(N_u-1)$ is the net potential energy injection by heat flux across the boundary; $N_E=F_E\Delta\phi I_0N_e/H$ is the net electric energy injection by current flux across the boundary; and $D_{elec} = F_Eq\langle\nabla\phi\cdot\nabla\phi-\frac{\partial \phi}{\partial t}\rangle_V
    +F_E\nu_E\langle\nabla\phi\cdot\nabla q\rangle_V$ is the electric dissipation~\cite{kourmatzis2012turbulent,guan2025numerical}. The domain-averaged kinetic energy, $\langle E_u\rangle_V$, and enstrophy, $\langle \Omega\rangle_V$, are shown in Table~\ref{tab:1}.

\subsection{Numerical analysis: numerical simulation, data-driven prediction, and coherent structure of energies and enstrophy with 2D wavelet decomposition and POD}
Direct numerical simulation (DNS) solves Eqs.~\eqref{eq:conMass}-~\eqref{eq:Temperature} using a Fourier-Chebyshev spectral solver at high spatial-temporal resolution, as described in our recent paper~\cite{guan2025numerical}. Specifically, the computational domain is $[x,y]\in[0:4,-1:1]$. The flow is periodic in the horizontal ($x$-) direction and wall-bounded (no-slip) in the vertical ($y$-) direction. The spatial derivative in the periodic horizontal direction is solved by the Fourier pseudo-spectral method, and the spatial derivative in the wall-normal direction is solved using the Chebyshev differential matrices~\cite[e.g.,][]{Trefethen2000spectral}. The temporal advancement is solved using the $2^{nd}$-order Adams–Bashforth method for the convection/advection terms, and the semi-implicit Crank–Nicolson method for the diffusion terms~\cite[]{leveque2007finite,guan2021stable}. The DNS timestep size is taken to be $\Delta t = 10^{-5}$ for all simulations. The vorticity ($\omega$) - stream function ($\psi$) formulation is used to represent the flow field, and the multiphysical convection is driven by heat transport, charge transport, and electric potential equations.

\subsubsection{Predicting the time series of domain-averaged energy equations using LSTM}
The dynamical system described by Eqs.~\eqref{eq:eu_vs_t}-\eqref{eq:ee_vs_t} represents the temporal evolutions of $\langle E_u\rangle_V$, $\langle E_p\rangle_V$, and $\langle E_e\rangle_V$. However, the algorithmic form of this system cannot be easily derived due to the complex relationships between the right-hand-side terms (e.g., $D_u$, $N_\theta$, etc) and the three left-hand-side variables (i.e., $\langle E_u\rangle_V$, $\langle E_p\rangle_V$, and $\langle E_e\rangle_V$). Here, we aim to test the prediction performance of the LSTM recurrent neural network in predicting the temporal evolutions of $\langle E_u\rangle_V$, $\langle E_p\rangle_V$, and $\langle E_e\rangle_V$ from previous timesteps of $\langle E_u\rangle_V$, $\langle E_p\rangle_V$, and $\langle E_e\rangle_V$ only. In this work, we apply a single-layer LSTM with a varying number of neurons ($n_{neuron}$) in the hidden layer (LSTM layer). We investigate the prediction accuracy over a range of look-back timesteps, $\Delta_\text{back}$ (the number of timesteps used for prediction), prediction horizon, $\Delta_\text{forward}$ (the number of timesteps forward predicted by the LSTM), and other hyperparameters such as batch size and $n_{neuron}$. Here, the data separating timestep $\Delta = 0.01 = 10^3\Delta t$.

\subsubsection{2D wavelet decomposition}
To prepare the data for data-driven model decomposition (i.e., POD), we first project the energy data (i.e., $E_u$, $E_p$, $E_e$) and the enstrophy data ($\Omega$) from the Chebyshev grid to a uniform grid in the wall-normal ($y$) direction~\cite{driscoll2014chebfun}. Next, we perform 2D wavelet decomposition and obtain the filtered snapshots (approximation) of $E_u$, $E_p$, $E_e$, and $\Omega$ at the second level such that the spatial resolution is reduced by 4 in each of the $x$ and $y$ directions~\cite{mallat1999wavelet}, which can be obtained using the {\it wavedec2} function in MATLAB~\cite{higham2016matlab}. The conceptual algorithm is described as follows, with $E_u(x,y)$ as an example.
\begin{eqnarray}
    E_u(x,y) &=& A_1+H_1+V_1+D_1,
\end{eqnarray}
where $A_1$ is the {\it approximation}, $H_1$ is the {\it horizontal detail}, $V_1$ is the {\it vertical detail}, and $D_1$ is the {\it diagonal detail} matrices all at the first level. The decomposition is performed using a low-pass filter and a high-pass filter, and in this work, we use the basic Haar wavelet (Haar filters). To further perform multiscale analysis, the {\it approximation} $A_1$ can be further decomposed in a similar way:
\begin{eqnarray}
    A_1 &=& A_2+H_2+V_2+D_2,\label{eq:A2}
\end{eqnarray}
where $A_2$, $H_2$, $V_2$, and $D_2$ are corresponding matrices at the second level~\cite{mallat1999wavelet}.

\subsubsection{Proper orthogonal decomposition (POD)}
To perform POD of the energy and enstrophy data, we first concatenate all snapshots of each of $E_u$, $E_p$, $E_e$, and $\Omega$ into a matrix $\mathbf{X}$, e.g., for $E_u$:
\begin{align}
\mathbf{X}=\begin{bmatrix}
E_u(0), & E_u(\Delta t), & E_u(2\Delta t), & \cdots & E_u(n_t\Delta t)\\
\end{bmatrix},
\end{align}
where $E_u(j\Delta t),\quad j=0,1,...,n_t$ is a column vector with a dimension reshaped from $[n_x,n_y]$ to $[n_x\times n_y,1]$, $n_t$ is the number of snapshots, $n_x$ and $n_y$ are the dimensions of the grid in the original two directions $x$ and $y$, and $\Delta t$ is the time difference in each two consecutive snapshots. POD of $E_u$ is obtained by performing a reduced singular value decomposition (reduced SVD) of $\mathbf{X}$: $[U,S,V^T] = svd(\mathbf{X}, ``econ")$~\cite{guan2021sparse,kutz2016dynamic,brunton2019data}. Here, matrix $U$ contains POD modes that contain spatial information (each column of $U$ is a reshaped POD mode), diagonal matrix $S$ contains the magnitude of each mode, and matrix $V$ contains their corresponding temporal evolution dynamics (each column of $V$ is a temporal evolution dynamic).

\section{Results} \label{sec:Results}
Table~\ref{tab:1} shows the domain and time averaged kinetic energy and enstrophy for the 8 cases. It can be observed that for constant parameters, the electric field has a positive impact on both kinetic energy and enstrophy, by comparing Cases 1 and 2 or Cases 5 and 6. However, when $F_\theta$ decreases, both the kinetic energy and the enstrophy decrease.

\begin{figure}[tbp!]
	\centering
	 \begin{overpic}[width=0.8\linewidth,height=0.8\linewidth]{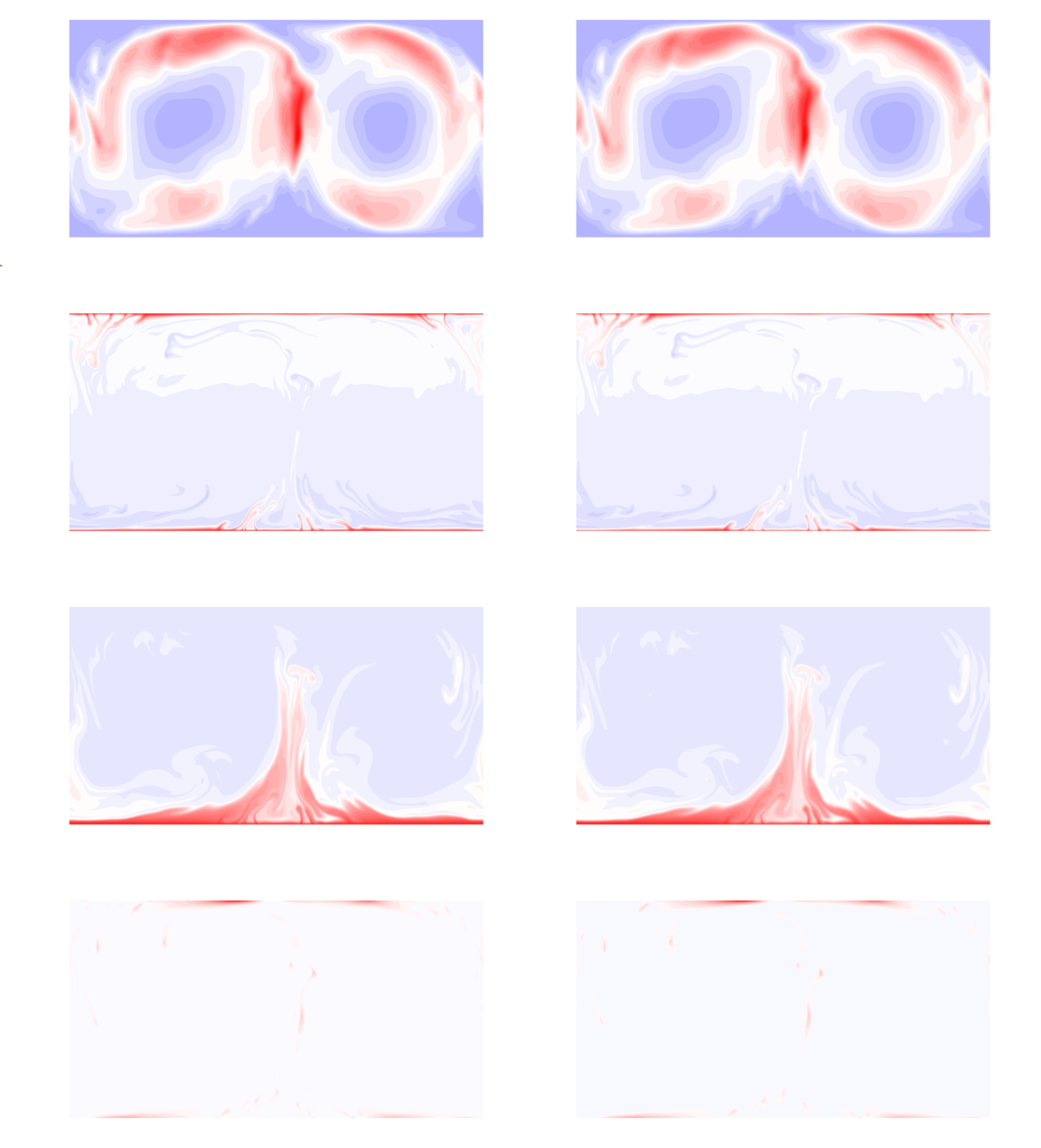}
     \put(22, 100) {DNS}
     \put(72, 100) {$A_2$}

     \put(-3,87){$E_u$}
     \put(-3,63){$E_p$}
     \put(-3,37){$E_e$}
     \put(-3,12){$\Omega$}
 \end{overpic}
	\caption{\small The DNS snapshots and their 2D wavelet approximations at the second level ($A_2$) of $E_u$, $E_p$,  $E_e$, and $\Omega$ for Case 8.}
    \label{fig:DNS_wavelet}
\end{figure}

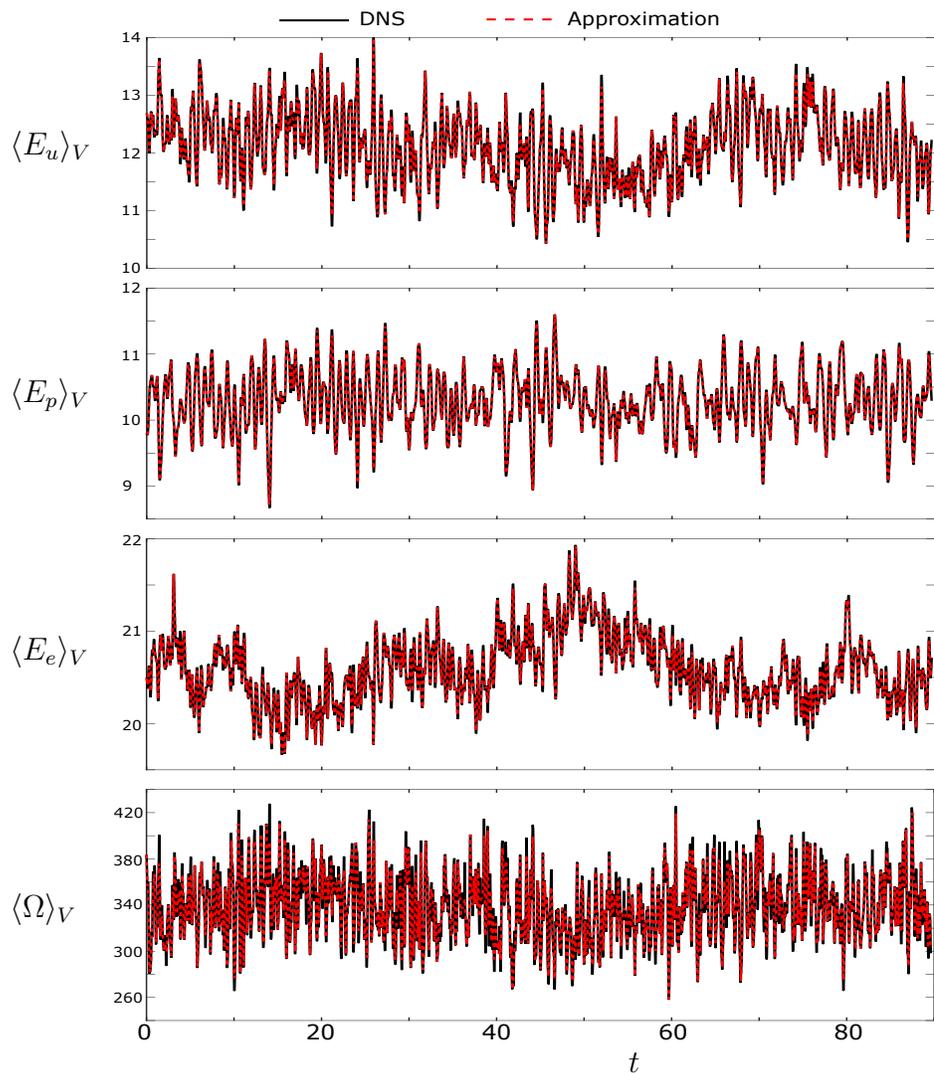
\begin{figure}[tbp!]
	\centering
	 \begin{overpic}[width=0.8\linewidth,height=1\linewidth]{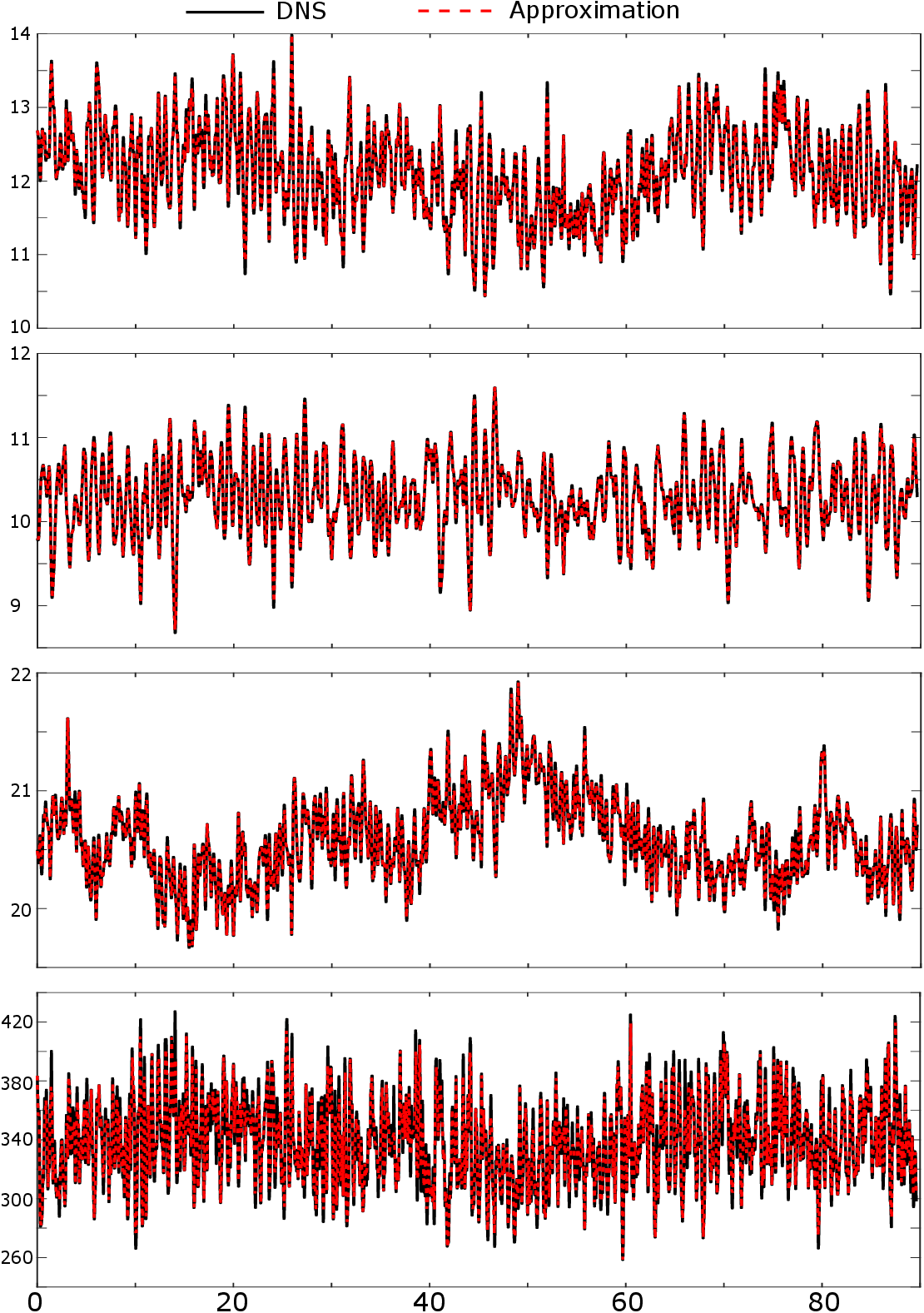}
     \put(-10,86){$\langle E_u\rangle_V$}
     \put(-10,62){$\langle E_p\rangle_V$}
     \put(-10,37){$\langle E_e\rangle_V$}
     \put(-10,12){$\langle \Omega\rangle_V$}

     \put(50,-3){$t$}
 \end{overpic}
	\caption{\small The domain-averaged DNS energy and enstrophy and their 2D wavelet approximations at the second level of $E_u$, $E_p$,  $E_e$, and $\Omega$ for Case 8. The domain-averaged values vary in time, $t$.}
    \label{fig:time_series_DNS}
\end{figure}

Next, we investigate the spatial and temporal distribution of the energy terms $E_u$, $E_p$, and $E_e$. Figure~\ref{fig:DNS_wavelet} shows the snapshots of $E_u$, $E_p$, and $E_e$ and their corresponding 2D wavelet approximation at the second level ($A_2$ as in Eq.~\eqref{eq:A2}). It can be observed that $A_2$ matches the DNS snapshots structurally, albeit using $16\times$ lower resolution. In fact, as shown in Fig.~\ref{fig:time_series_DNS}, the domain-average values $\langle E_u\rangle_V$, $\langle E_p\rangle_V$, and $\langle E_e\rangle_V$ obtained from the approximation $A_2$ also match the DNS well, meaning that the energies are primarily contained in the large scales. It further justifies our choice of using 2D wavelet decomposition to approximate the energy distribution in 2D ETHD systems.

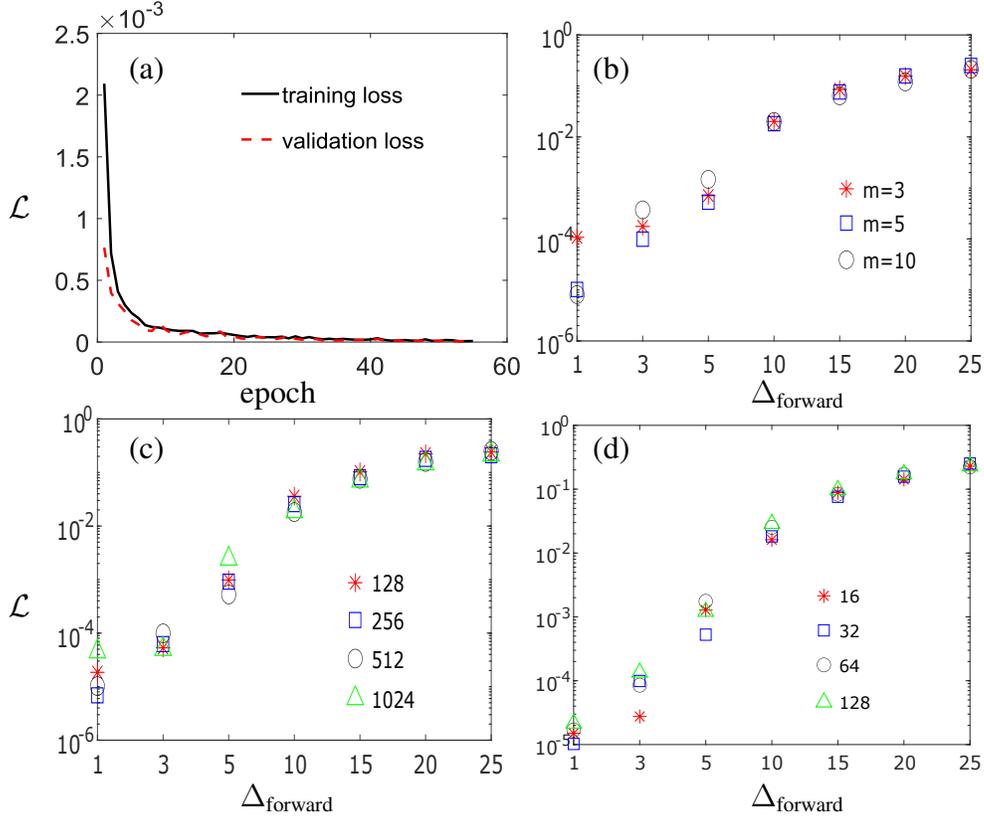
\begin{figure}[tbp!]
	\centering
	 \begin{overpic}[width=0.9\linewidth,height=0.75\linewidth]{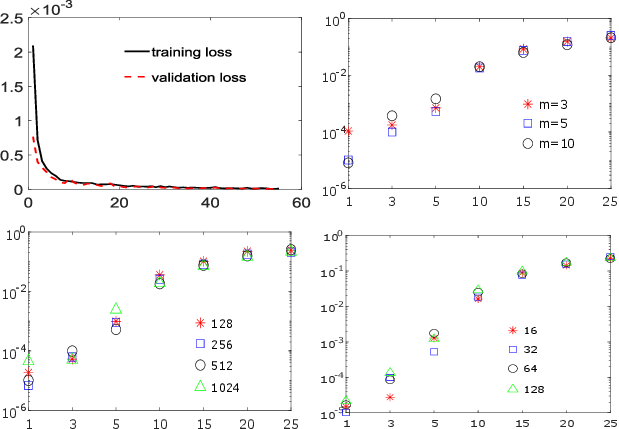}
     \put(8, 75) {(a)}
     \put(58, 75) {(b)}
     \put(8, 34) {(c)}
     \put(58, 34) {(d)}

     \put(-5,60){$\mathcal{L}$}
     \put(-5,17){$\mathcal{L}$}

     \put(20,40){epoch}
     \put(75,40){$\Delta_\text{forward}$}
     \put(20,-3.5){$\Delta_\text{forward}$}
     \put(75,-3.5){$\Delta_\text{forward}$}

 \end{overpic}
 \vspace{2mm}
	\caption{\small (a) The training and validation loss $\mathcal{L}$ for Case 8 with $\Delta_\text{forward}=\Delta$, $\Delta_\text{back}=10\Delta_\text{forward}$, $n_{neuron} = 256$, and batch size $=32$; (b) the minimum validation loss ($min(\mathcal{L})$) with respect to $\Delta_\text{forward}$ and $\Delta_\text{back}=m\Delta_\text{forward}$ with $n_{neuron} = 512$, and batch size $=16$; (c) $min(\mathcal{L})$ with respect to $\Delta_\text{forward}$ and $n_{neuron} = [128, 256, 512, 1024]$ with $\Delta_\text{back}=5\Delta_\text{forward}$ and bath size $=16$; and (d) $min(\mathcal{L})$ with respect to $\Delta_\text{forward}$ and batch size $= [16, 32, 64, 128]$ with $\Delta_\text{back}=5\Delta_\text{forward}$ and $n_{neuron}=512$.}
    \label{fig:loss}
\end{figure}

\begin{figure}[tbp!]
	\centering
	 \begin{overpic}[width=1.0\linewidth,height=0.8\linewidth]{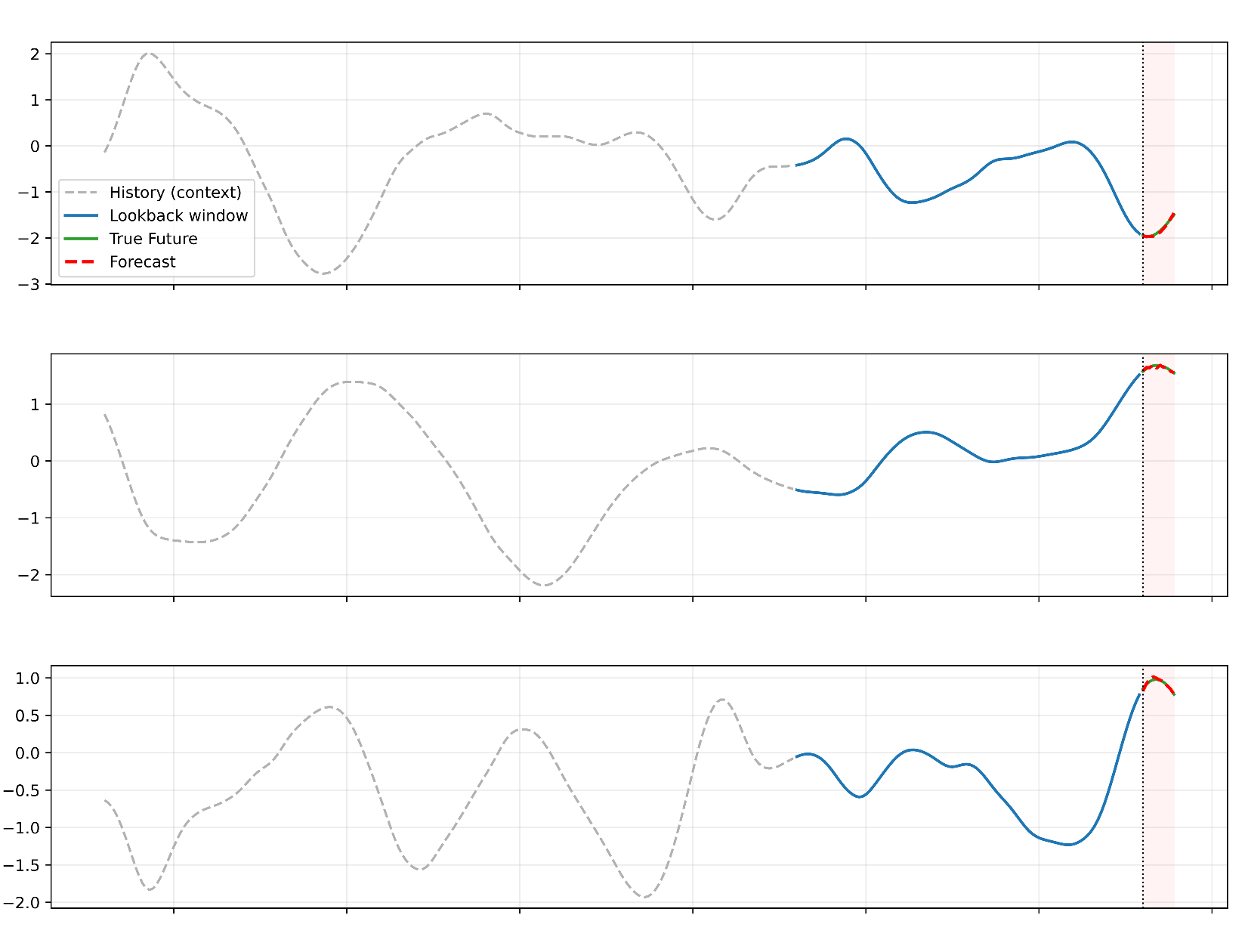}
     \put(45, 77.5) {$\langle E_u\rangle_V/\langle E_u\rangle_{V,t}$}
     \put(45, 52) {$\langle E_p\rangle_V/\langle E_p\rangle_{V,t}$}
     \put(45, 25.5) {$\langle E_e\rangle_V/\langle E_e\rangle_{V,t}$}
     
     \put(52,-3){$t$}

     \put(82, -0.5){$89$}
     \put(54, -0.5){$88$}
     \put(27, -0.5){$87$}
 \end{overpic}
 \vspace{2mm}
	\caption{\small Prediction of the normalized time series $\langle E_u\rangle_V$, $\langle E_p\rangle_V$, and $\langle E_e\rangle_V$.The LSTM has $\Delta_\text{forward} = 10\Delta$, $\Delta_\text{back} = 10\Delta_\text{forward}$, $n_{neuron}=512$, and batch size $=32$.}
    \label{fig:LSTM}
\end{figure}

As we have the temporal evolution of $\langle E_u\rangle_V$, $\langle E_p\rangle_V$, and $\langle E_e\rangle_V$ representing the dynamical system mathematically described in Eqs.~\eqref{eq:eu_vs_t}, \eqref{eq:ep_vs_t}, and \eqref{eq:ee_vs_t}, we investigate the performance of a single-layer LSTM in prediction. Here, we use the time series shown in Fig.~\ref{fig:time_series_DNS} normalized by their corresponding temporal average (i.e., $\langle E_u\rangle_V/\langle E_u\rangle_{V,t}$, $\langle E_p\rangle_V/\langle E_p\rangle_{V,t}$, and $\langle E_e\rangle_V/\langle E_e\rangle_{V,t}$) as training and validation data. Here, we use the time series in $t\in[0,80]$ as training data set, $t\in(80,85]$ as validation data set, and $t\in(85,89.4]$ as testing data set. Figure~\ref{fig:loss} shows the training processes from a variety of single-layer LSTMs with different hyperparameters $\Delta_\text{forward}$, $\Delta_\text{back}$, $n_{neuron}$, and batch size. Figure~\ref{fig:loss} (a) shows the training loss and validation loss over epochs (training iterations). In most cases, the training is converged within 60 epochs.  Figure~\ref{fig:loss} (b) shows the effects of the prediction horizon $\Delta_\text{forward}$ and the number of look-back timesteps $\Delta_\text{back}$. It can be observed that a smaller $\Delta_\text{forward}$ leads to higher prediction accuracy (lower $\mathcal{L}$), which is intuitive because the closer the future is, the easier it is to predict. However, $\Delta_\text{back}$ does not follow ``the more the better'' rule. Here, we use $\Delta_\text{back}=m\Delta_\text{forward}$, and find that $m=5$ actually performs better than $m=3$ or $m=10$ in terms of reducing the validation loss. Figure~\ref{fig:loss} (c) shows the effects of $n_{neuron}$, and we find that $n_{neuron}=512$ works well in all cases. A small $n_{neuron}$ may lead to underfitting and a large $n_{neuron}$ may lead to overfitting, which both will deteriorate the validation loss. Figure~\ref{fig:loss} (d) shows the effects of batch size, and we find that batch size $=16$ or $32$ works well for all cases. 
In general, to balance the prediction horizon and accuracy, we use $\Delta_\text{forward} = 10$, $\Delta_\text{back} = 10\Delta_\text{forward}$, $n_{neuron}=512$, and batch size $=32$ for the next analysis, which is shown in Fig.~\ref{fig:LSTM}.

The LSTM testing results are shown in Fig.~\ref{fig:LSTM}. The testing data set has not been seen during the training process, as the LSTM is trained on the training data, and the stopping criteria and the optimization are based on the validation data. The testing data set is used for testing the prediction performance only. It can be seen from Fig.~\ref{fig:LSTM} that the prediction (forecast) matches the truth (true future) very well using the LSTM in the prediction horizon ($10\Delta)$, where $\Delta=10^3\Delta t$ is the data separating timestep. It should be noted that the prediction horizon shown in Fig.~\ref{fig:LSTM} includes an extreme value (local minimum in $\langle E_u\rangle$ or maximum in $\langle E_p\rangle$ and $\langle E_e\rangle$), which shows the capability of the LSTM to predict not only monotonic behavior but also to correctly predict extreme values in a chaotic time series. Although the prediction horizon seems to be short compared with the full time series, it should be noted that the prediction horizon here is equivalent to $10^4\Delta t$, where $\Delta t$ is the DNS timestep size.

\begin{figure}[tbp!]
	\centering
	 \begin{overpic}[width=0.8\linewidth,height=0.6\linewidth]{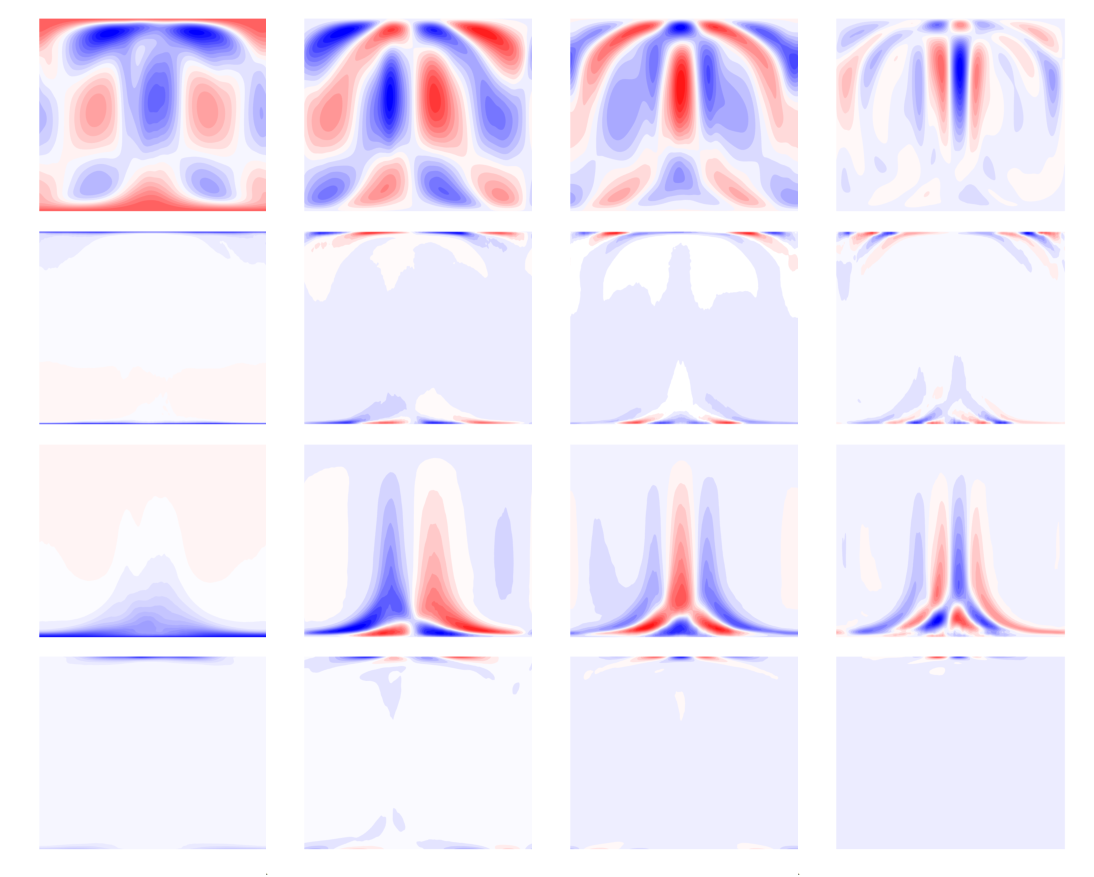}
     \put(8, 75) {mode 1}
     \put(33, 75) {mode 2}
     \put(56, 75) {mode 3}
     \put(82, 75) {mode 4}

     \put(-3,64){$E_u$}
     \put(-3,45){$E_p$}
     \put(-3,27){$E_e$}
     \put(-3,10){$\Omega$}
 \end{overpic}
	\caption{\small The first 4 POD modes of $E_u$, $E_p$,  $E_e$, and $\Omega$ for Case 8.}
    \label{fig:POD_modes}
\end{figure}
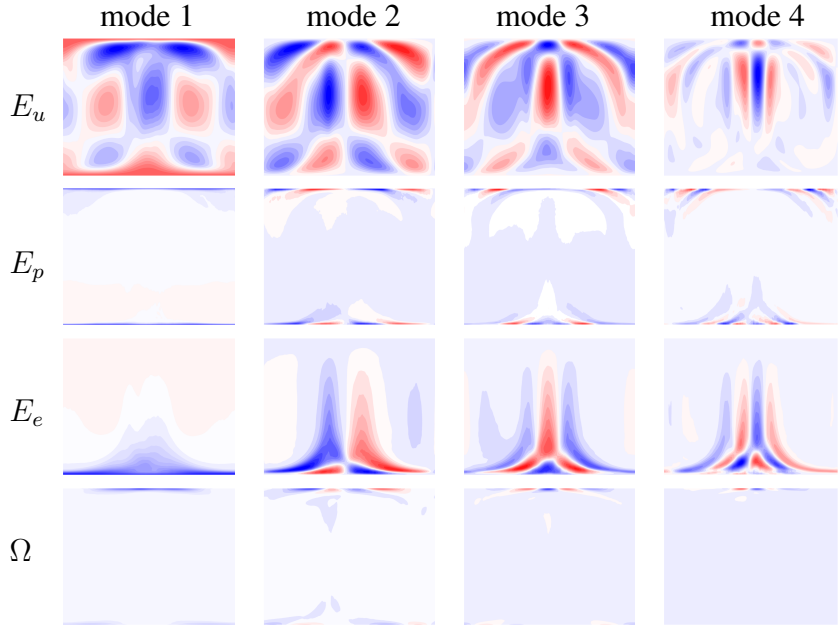

The temporal evolution of domain-average values shows how energies vary over time. However, the spatial distribution information is lost when the domain average is applied. Next, we investigate the spatial distribution of the energy and enstrophy terms by showing the results of POD modal analysis of $E_u$, $\Omega$, $E_p$, and $E_E$ of Case 8, as shown in Fig.~\ref{fig:POD_modes}. It can be observed that for all the energy-related properties, the first POD mode captures the large-scale dynamics (coherent structures) of the corresponding energy. For kinetic energy $E_u$, the first three modes contains energy in the large scales. For $E_p$, $E_e$, and $\Omega$, small-scale structures (with smaller wave length in $x$) starts to emerge from mode 2.  Also, the potential energy is more evenly distributed within the whole domain, while the electric energy is more concentrated in the plume. The findings here are consistent with our previous results on $\Phi_\theta$ and $\Phi_E$~\cite{guan2025numerical}. It should also be noted that $\Omega$ is heavily distributed within the boundary layer near the upper and lower walls. This is because in 2D flow field, enstrophy $\Omega$ is directly related to the viscous dissipation which is maximum near the wall due to the high velocity gradient~\cite{grossmann2000scaling, davidson2015turbulence,guan2025numerical}. 

\begin{figure}[tbp!]
	\centering
	 \begin{overpic}[width=0.8\linewidth,height=0.8\linewidth]{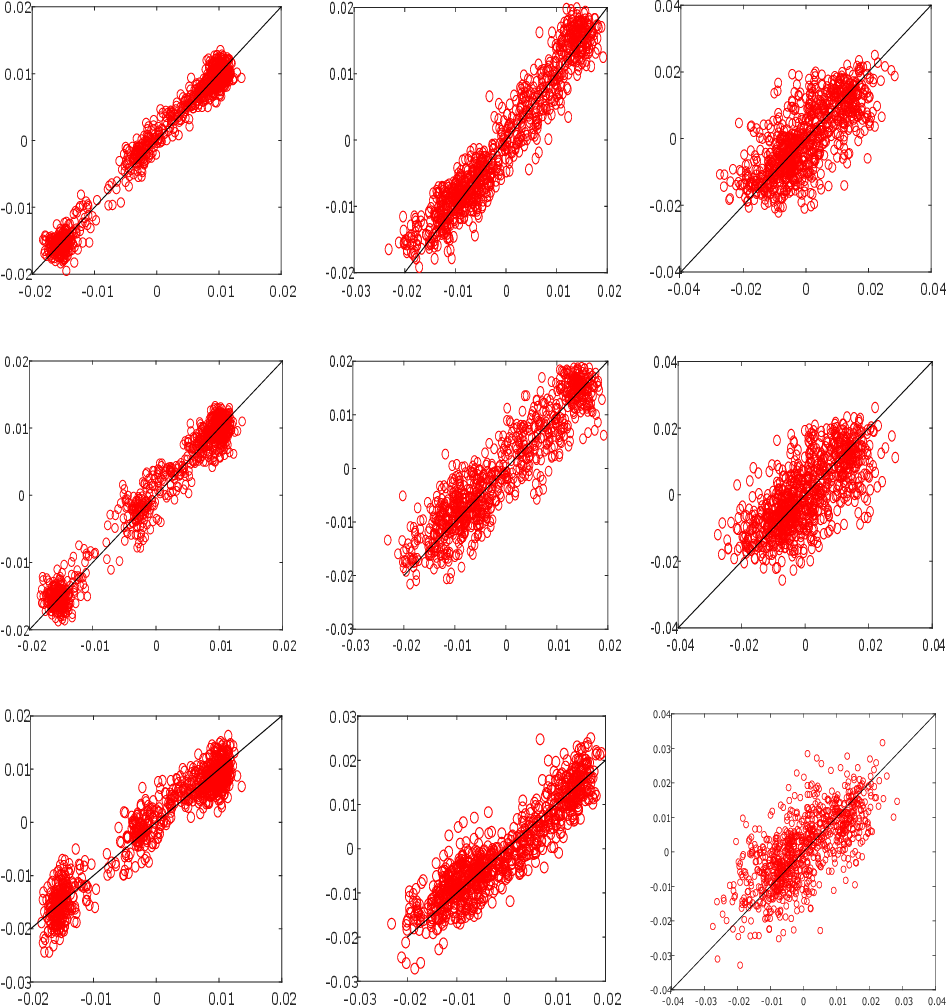}
     \put(12, 100) {mode 2}
     \put(45, 100) {mode 3}
     \put(80, 100) {mode 4}

     \put(13,67.5){\footnotesize $E_u$}
     \put(46,67.5){\footnotesize $E_u$}
     \put(81,67.5){\footnotesize $E_u$}
     
     \put(13,32){\footnotesize $E_u$}
     \put(46,32){\footnotesize $E_u$}
     \put(81,32){\footnotesize $E_u$}
     
     \put(13,-4){\footnotesize $E_u$}
     \put(46,-4){\footnotesize $E_u$}
     \put(81,-4){\footnotesize $E_u$}

     \put(-4,85){\footnotesize $E_p$}
     \put(-4,49){\footnotesize $E_e$}
     \put(-4,13){\footnotesize $\Omega$}
 \end{overpic}
  \vspace{2mm}
	\caption{\small The correlation between the $2^{nd}$, $3^{rd}$, and $4^{th}$ POD mode coefficients (columns of the matrix $V$) of $E_p$,  $E_e$, and $\Omega$ with  $E_u$ for Case 8. The black lines represent the linear fit.}
    \label{fig:POD_coeffs}
\end{figure}
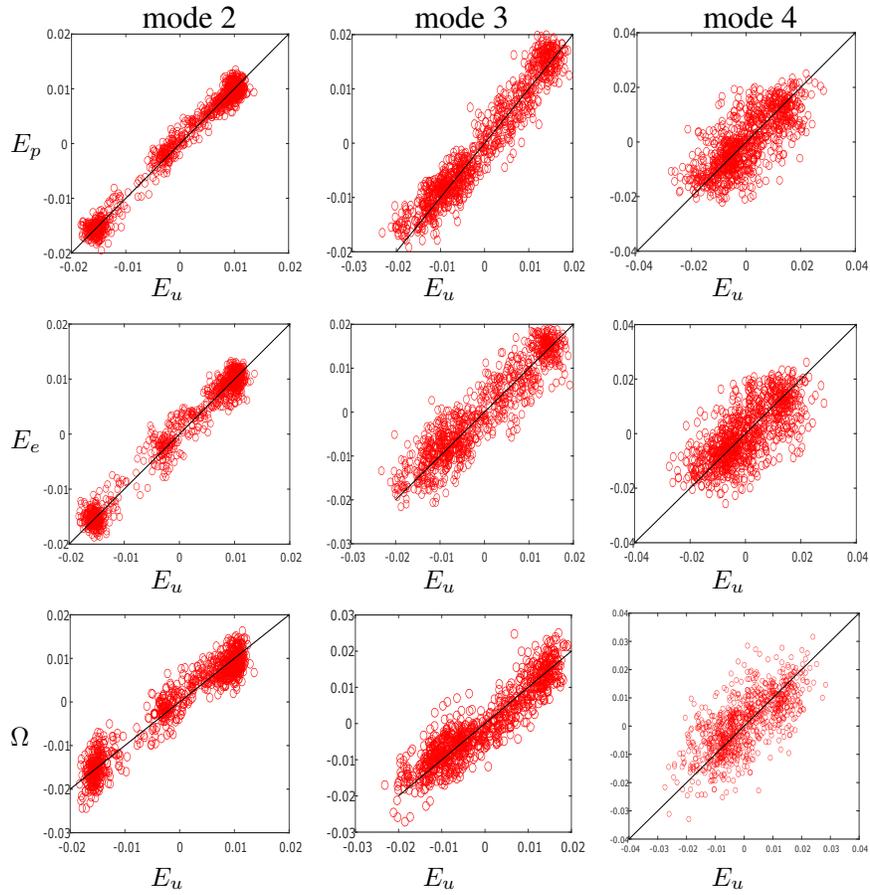

In addition to the spatial distribution of energy and enstrophy, POD analysis also demonstrates the correlation between different modes obtained from different quantities. Figure~\ref{fig:POD_coeffs} shows the correlation between the modal coefficients of $E_p$,  $E_e$, and $\Omega$ with  $E_u$. It can be observed that these modal coefficients are highly correlated (with correlation coefficients above $0.9$), forming an almost linear relationship. This is particularly interesting, as it suggests that given a few leading modes, $E_p$,  $E_e$, and $\Omega$ can be reconstructed by knowing or correctly predicting the coefficients of $E_u$ modes only.

\section{Conclusion}\label{sec:conclusion}
In this work, we analyze the energy terms in 2D ETHD turbulent convection systems. We first derive the governing equations (as in a dynamical system) of the kinetic energy, potential energy, and electric energy. We then remark the transitional terms ($\Phi_\theta$ and $\Phi_E$), which are the different contributions to kinetic energy due to potential energy and electric energy. This analysis completes the energy transfer picture described earlier in our previous work~\cite{guan2025numerical}, and it is firstly reported in this work. 

Next, we investigate a variety of single-layer LSTMs in predicting the time series of domain-average kinetic energy, potential energy, and electric energy. We show the capability of LSTM to predict up to $10^4\Delta t$ accurately and predict the local extreme values accurately.

Furthermore, we perform POD analysis on the energy and enstrophy terms and show that the coherent structure that contains the energy can be well described by the first POD mode. We also investigate the correlation of modal coefficients between different energy/enstrophy terms and find that they form a linear relationship. However, due to the same feature, POD analysis cannot reveal the information contained in small structures which may be important for turbulent modeling using e.g., large eddy simulations.

In our future work, we aim to extend our modal analysis by using different data-driven modal analysis, such as dynamic mode decomposition (DMD), empirical mode decomposition (EMD), and variational mode decomposition (VMD), and compare their performance on analyzing the energy in a 2D ETHD turbulence system.

\section{Acknowledgment}\label{sec:acknowledgment}
The authors appreciate the discussion with Dr. Mengqi Zhang, Dr. Qi Wang, and Dr. Yu Zhang on the chaotic ETHD systems.

\bibliographystyle{unsrt}
\bibliography{ISEHD_JE.bib}

\end{document}